\documentclass[9pt,twocolumn,twoside]{opticajnl}
\journal{opticajournal} 

\setboolean{shortarticle}{false}

\usepackage{lineno}
\usepackage{braket}
\usepackage{amsmath}

\title{Polarization coverage and self-healing characteristics of Poincar\'{e}-Bessel beam}

\author[1,2]{Subith Kumar}
\author[1]{Anupam Pal}
\author[3]{Arash Shiri}
\author[1]{G. K. Samanta}
\author[3]{Greg Gbur}

\affil[1]{Physical Research Laboratory, Ahmedabad, Gujarat  380009, India}
\affil[2]{Indian Institute of Technology Gandhinagar, Ahmedabad, Gujarat 382424, India}
\affil[3]{Department of Physics and Optical Science, University of North Carolina Charlotte, Charlotte, North Carolina 28223}

\affil[*]{subithpkumar@gmail.com,, gsamanta@prl.res.in, gjgbur@uncc.edu}

\begin{abstract}
As a vector version of scalar Bessel beams, Poincar\'{e}-Bessel beams (PBBs) have attracted a great deal of attention due to the presence of polarization singularities and their nondiffraction and self-healing properties. Previous studies of PBBs have been restricted primarily to understanding the disinclination patterns in the spatially variable polarization, and many of the properties of PBBs remain unexplored. Here, we present a theoretical and experimental study of the polarization characteristics of PBBs, investigating a variety of their features. Using a mode transformation of a full Poincar\'{e} (FP) beam in a rectangular basis, ideally carrying 100$\%$ polarization coverage of polarization states represented on the surface of the Poincar\'{e} sphere, we observe the PBB as the superposition of an infinite number of FP beams, as each ring of PBB has polarization coverage >75$\%$. We also observe the resilience of a PBB's degree of polarization to perturbation. The polarization-ellipse orientation map of PBBs shows the presence of infinite series of C-point singularity pairs. The number of such series pairs is decided by the number of C-point singularity pairs of the FP beam. The dynamics of C-point singularity pairs in the self-healing process show a non-trivial creation of new singularities and recombination of existing singularities. Such dynamics provide insight into ``Hilbert Hotel'' style evolution of singularities in light beams. The present study can be useful for imaging in the presence of depolarizing surroundings, studying turbulent atmospheric channels, and exploring the rich mathematical concepts of transfinite numbers.
\end{abstract}

\setboolean{displaycopyright}{false} 

\begin{document}

\maketitle

\section{Introduction}

Beams with nontrivial tailoring of their spatial profile, now called spatially structured optical beams or structured light, have attracted a great deal of attention recently due to their wide range of applications. These include optical manipulation \cite{Otte2020OpticalManipulation, Wang2012OpticalBeams}, micromachining \cite{liu2018nonlinear, Toyoda2013TransferNanostructures, Omatsu2019ALight}, imaging \cite{Chen2013ImagingSystem, Torok2004TheMicroscopy}, and optical communications \cite{Ndagano2018CreationCommunication, Lei2017ApproachBeams}. Typically, structured beams are generated through mode conversion of a Gaussian beam, and can include modulation of the amplitude, phase, or polarization of the beam.  A beam with a uniform state of polarization is referred to as a scalar beam and a beam with a nonuniform state of polarization is referred to as a vector beam.

Perhaps the most familiar class of structured beams are scalar optical vortex beams, which possess an optical vortex on their central axis. An optical vortex is a line of zero intensity around which the phase has a circulating or helical structure; the study of these and related singularities now form a field known as singular optics \cite{Gbur2017SingularOptics}. In a transverse plane, an optical vortex manifests as a point, and the phase always changes by an integer multiple of $2\pi$ around the vortex; this multiple is called the topological charge. The topological charge is generally a conserved quantity and is typically only created or annihilated in pairs of equal and opposite charges. Vector vortex beams may also be generated through the superposition of vortex beams of different orders in orthogonal polarization states. The full Poincar\'{e} beam is a special class of vector vortex beams, ideally containing all polarization states that are present on the surface of the Poincar\'{e} sphere. Such beams have found practical application; for example, they have lower scintillation than comparable beams of uniform polarization in the presence of atmospheric turbulence \cite{Gu:09, Wei:15}. Recently, efforts have been made to find a general method to estimate the polarization coverage and the parameters influencing the polarization coverage of the Poincar\'{e} beam \cite{Kumar:23} to broaden the scope of such beams for different applications.

Bessel beams \cite{Khonina2020BesselReview, mcgloin2005bessel} are a different class of structured light beams that have found application in various areas of optics \cite{stoian2018ultrafast, liu2020simultaneous, dudutis2016non, ambrosio2011integral}, owing to their high intensity extended focus, finite beam width, nondiffractive propagation over considerable distances, and self-healing properties behind obstacles. Recently, there has been much interest in extending the scope of Bessel beams  through the generation of vector Bessel beams \cite{glukhova2022vector, bouchal1995non}, for use in studying quantum effects \cite{PhysRevA.71.033411, mclaren2014self}, microscopy, imaging \cite{fahrbach2010microscopy, planchon2011rapid}, and propagation in turbulent media \cite{Wei:15, Gu:09, cheng2016propagation}. Furthermore, efforts have been made to include the ideally 100$\%$ polarization coverage of the Poincar\'{e} beam to the Bessel beam through the generation of Poincar\'{e}-Bessel beams \cite {Shvedov:15, Holmes:19}, and study their structure and propagation characteristics. However, many interesting properties, such as the dynamics in the polarization pattern, changes in the degree of polarization, and polarization coverage of the beam, especially during the self-healing process, are left unanswered. 

In this paper, we report a theoretical and experimental study of the polarization characteristics of the Poincar\'{e}-Bessel beam. Using an axicon, we have transformed a Poincar\'{e} beam in the rectangular basis having polarization coverage >98$\%$ into a Poincar\'{e}-Bessel beam. The study of the polarization characteristics of the Poincar\'{e}-Bessel beam reveals important features, such as each ring of the Poincar\'{e}-Bessel beam behaving like the Poincar\'{e} beam with polarization coverage >75$\%$, self-healing of the polarization structure, and degree-of-polarization is independent of the beam obstruction. Using the polarization ellipse orientation map having an infinite number of C-point singularity pairs in the self-healing process, we study the dynamics of the C-point singularities in the reconstruction process and consider its connections to the mathematics of infinite sets.

\section{Theory}
To verify our experimental results, we have analytically investigated the evolution of polarization singularities of the Poincar\'{e}-Bessel beam due to diffraction effects. First, the  physics and mathematics of Poincar\'{e}-Bessel beams are briefly introduced and then its propagation in the presence of an obstacle is studied semi-analytically with the aid of numerical methods.
\subsection{Poincar\'{e} beam and polarization singularities}
In a transverse electromagnetic field, the polarization is defined as the direction of the electric field which is determined by the ratio of amplitudes of the orthogonal components of the field in the transverse plane ($E_x$ , $E_y$ in x-y basis). Generally, these two components of E-field do not oscillate in phase with each other and as the beam propagates, the field vector traces out an ellipse on the transverse plane. Determining the polarization state of a beam is equivalent to finding the geometric properties of its corresponding ellipse. Polarization singularities are the singularities of characterizing the direction (orientation and helicity) of the polarization ellipse. The orientation angle $\Psi$ is defined as the angle between the major axis of the ellipse and the positive direction of $x$ axis. C-points are the singularity points of circular polarization in which the major axis and hence the orientation angle is undefined. L-lines are the singularity lines of the linearly polarized beam on which the helicity (handedness) of the polarization ellipse is undefined.\\ However, these geometrical features of light are not easily measurable. In 1852, Stokes introduced four experimentally measurable parameters which can be explicitly defined in terms of both the E-field and the polarization ellipse \cite{ggs:tcps:1852}
\begin{equation}\label{t1}
    \begin{split}
        S_0 &= E_x^2 + E_y^2 ,\\
        S_1 &= E_x^2 - E_y^2 , \\
        S_2 &= 2|E_x||E_y| cos\delta ,\\
        S_3 &= 2|E_x||E_y| sin\delta .
    \end{split}
\end{equation}
where $\delta$ is the phase difference between $E_x$ and $E_y$. Spanning a 3-D space by the unit Stokes parameters $(\hat{s_1},\hat{s_2},\hat{s_3})$, we can build up a unit sphere known as Poincar\'{e} sphere which provides us a complete pictorial illustration of the polarization state of light\cite{}. Any point on the surface of the Poincar\'{e} sphere represents a distinct completely polarized state. North and south poles correspond to the right and left-handed circular polarization, respectively (C-points), and all points on the equator of the sphere represent linear polarization states (L-lines).\par
Optical vector beams with spatially inhomogeneous distribution of polarization vector on the transverse plane are called Poincar\'{e} beams. In a Poincar\'{e} beam, the polarization distribution occupies a number of points on the Poincar\'{e} sphere and a full Poincar\'{e} beam takes up the entire surface of the sphere.\cite{Beckley:10} \\
\subsection{Poincar\'{e}-Bessel beams}
Non-diffracting Bessel beams were first introduced in 1987 by Durnin as the precise solution of the scalar wave equation.\cite{jd:josaa:1987}
In cylindrical coordinates, the complex amplitude of the beam contains an $l^{th}$ order Bessel function known as Bessel beam of mode $l$
\begin{equation}\label{t2}
    U_l(r,\phi,z) = \exp{(k_z z)}J_l(k_r r)e^{\pm il\phi}
\end{equation}
where $k_r , k_z$ are transverse and longitudinal components of the wave vector $k=2\pi / \lambda$, respectively. $\phi$ and $r$ are the azimuthal angle and radial distance from the axis on the cross-section of the beam, respectively. $J_l$ is the $l^{th}$ order Bessel function of the first kind.\par
Poincar\'{e}-Bessel beams are prepared by vector superposition of coherent Bessel modes in orthogonal coordinate bases.
In our experiment, we have used the superposition of orthogonal $1^{st}$ and $0^{th}$ order of the Bessel beam in Cartesian coordinate as the source.
\begin{equation}\label{t3}
     \mathbf{E_0(\mathbf{r})} = \frac{1}{\sqrt{2}}\left[J_1(k_r r) e^{\pm i\phi} \hat{x} + J_0(k_r r)\hat{y}\right].
\end{equation}
The disclinations are characterized by the orientation of the polarization ellipse $\Psi$ and the index of polarization singularities is defined as\cite{nye1974dislocations}
\begin{equation}
I_c = \frac{1}{2\pi}\oint _C d\Psi \; ,
\end{equation}
the integration is over a closed path $C$ encircling the singularities. $I_C$ is an integer value implying that the continuous function $\Psi(\mathbf{r})$ will have a discontinuous jump as it runs into a singularity.  Depending on the sign of the topological index, the lemon or star disclination will appear.\cite{Gbur2017SingularOptics} \par
The combination of the non-diffracting self-healing character of Bessel beams with the spatially varying polarization of Poincar\'{e} beams exhibits interesting diffraction behavior in propagation through an aperture or around an obstacle.
\subsection{Diffraction by passing an obstacle}
\begin{figure}[htp!]
 \centering
 \includegraphics[scale=0.5]{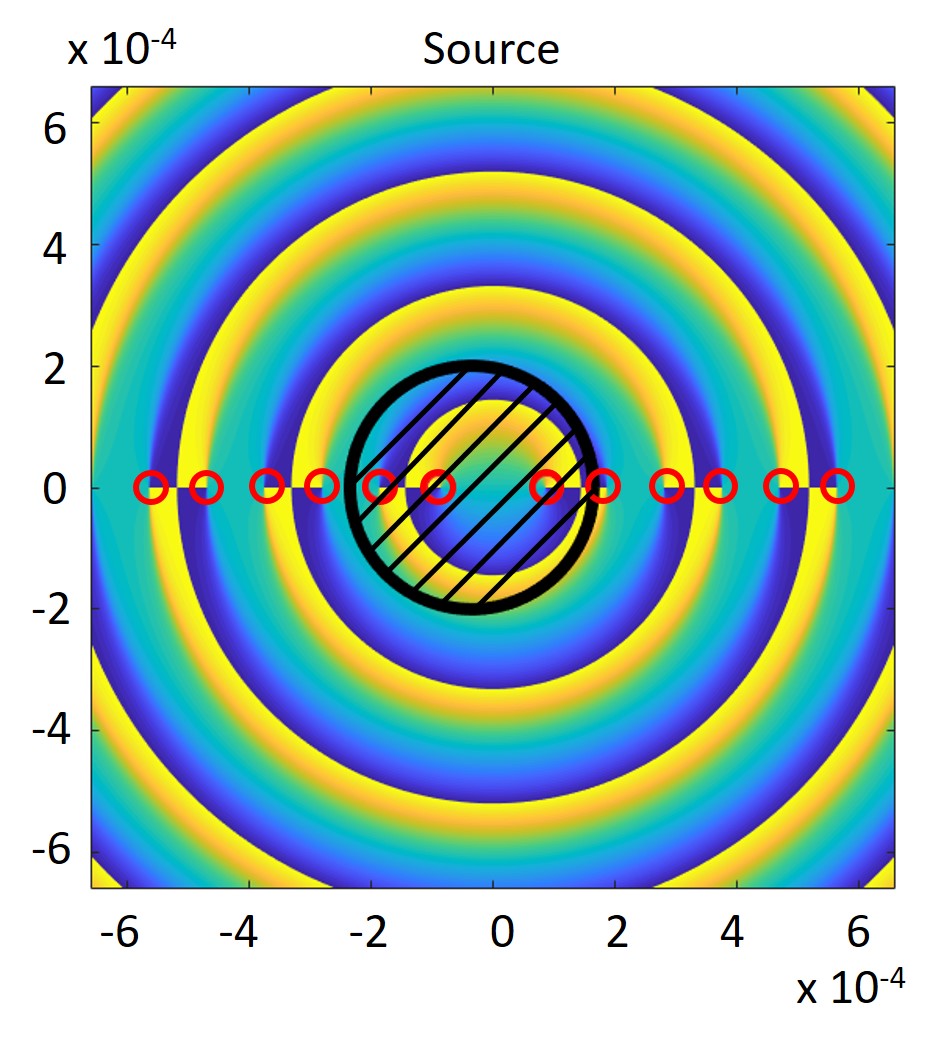}
  \caption{Stokes vector phase of the obstructed Poincar\'{e}-Bessel beam of Eq. (\ref{t3}). The solid black circle shows the extent of the region covered by the spherical obstruction. The singularities blocked by the sphere are marked by red circles with the total topological index shifted by $\pm 1$.}
  \label{turbulencefig_1}
\end{figure}

To investigate the self-healing effects on the evolution of polarization singularities, we consider the propagation of Poincar\'{e}-Bessel beam expressed in  Eq. (\ref{t3}) obstructed by an off-axis sphere defined as
\begin{equation}\label{t4}
    circ(r) = 
   \begin{cases}
1      \;\;\;\;\;\;    |r-r_0| > a\\
0   \;\;\;\;\;\;   |r-r_0| \leq a \;\;\;\;,
\end{cases} 
\end{equation}
where $a$ is the radius of the sphere, $r$ is the distance from the center in the transverse plane, and $r_0$ is the radial distance between the center of the block and the center of the beam. The values of $a$ and $r_0$ are chosen such that an odd number of singularities are blocked by the obstacle (see the black shaded region of Fig. \ref{turbulencefig_1}): two lemons and one star by which the net topological index is shifted by one unit of charge.\par
The Fresnel propagation of the beam perturbed by the block can be expressed as the integral over the shifted position vector $\mathbf{R} = \mathbf{r} - \mathbf{r}_0$ on the cross-section of the beam at the source as,
\begin{equation}
    \mathbf{E}(\mathbf{r}^\prime,z)=\frac{e^{ik_z z}}{i\lambda z}\iint \mathbf{E_0}(\mathbf{R}+\mathbf{r}_0) \exp\left[\frac{ik}{2z}\left|\mathbf{r}' - (\mathbf{R}+\mathbf{r}_0)\right|^2\right] d^2R\, ,
\end{equation}
where $\lambda$ is the wavelength of the beam and $\mathbf{r}$ and $\mathbf{r}'$ are the position vectors on the source and detector planes, respectively. Due to the interrelationship between the azimuthal angles of the position vectors $\mathbf{r}$, $\mathbf{r}_0$ and $\mathbf{R}$, along with the finite limits of the radial integral constrained by the radius of the spherical obstacle, the analytical calculation of the above Fresnel propagation is not possible. For this reason, we employ detailed numerical methods to show how the singularities of the beam recast after being disturbed by the obstruction and explore how the distinctive healing properties of the Bessel beam can exhibit a phenomenon similar to the Hilbert Hotel evolution. 
\begin{figure}[htp!]
 \centering
 \includegraphics[scale=0.4]{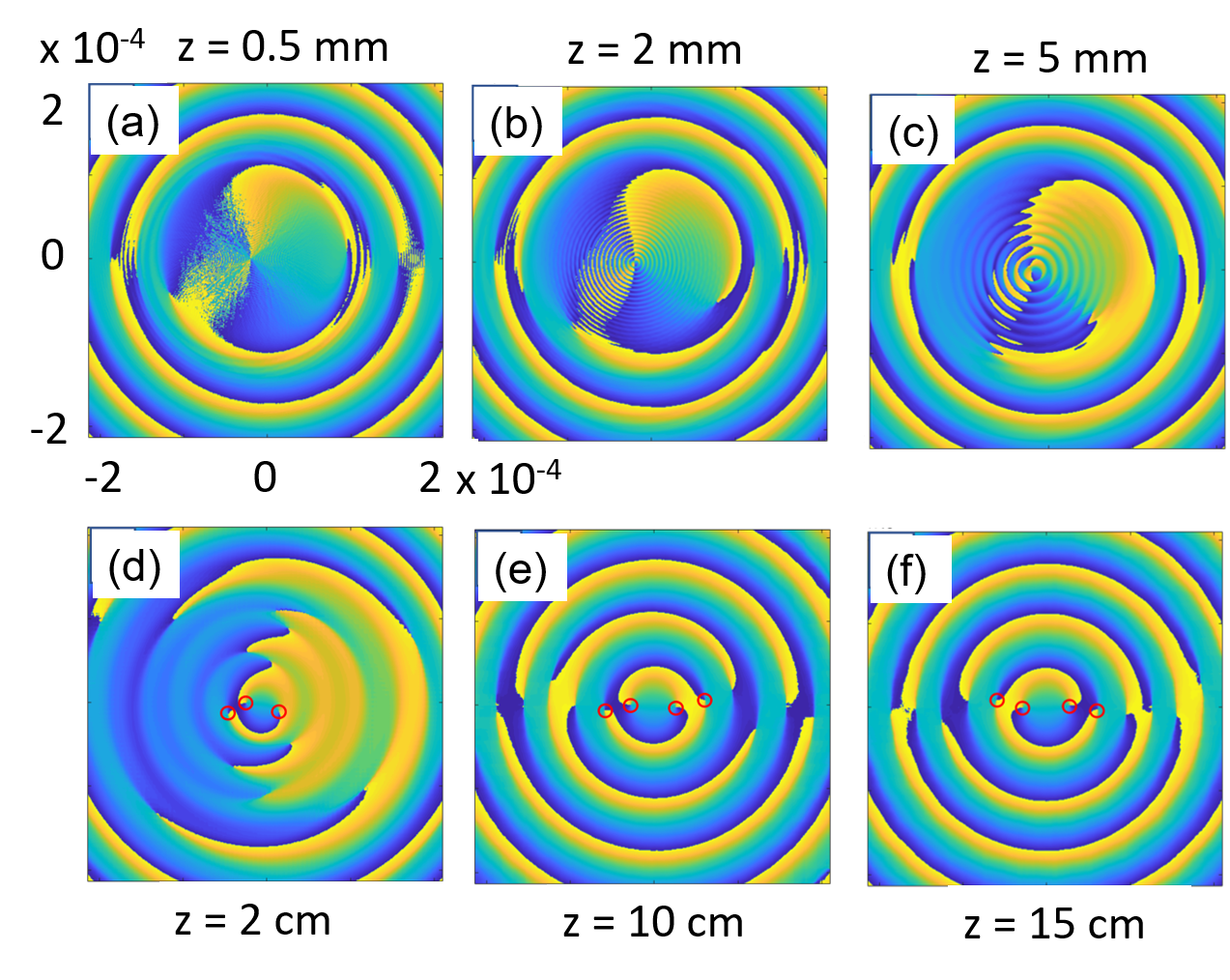}
 \caption{Hilbert's Hotels like evolution of polarization singularities in the self-healing process of the Poincar\'{e}-Bessel beam after passing an obstruction.}
  \label{turbulencefig_2}
\end{figure}
Figure \ref{turbulencefig_2} shows the Stokes vector phase (orientation angle) of the beam in different propagation distances from $0.5\,$ mm to $15\,$ cm after the block. As shown in Fig. \ref{turbulencefig_1}, two lemons and one star singularities were blocked by the obstruction leaving an unbalanced topological index in the inner rings. Due to the disturbing effects of the block, as evident from Fig. \ref{turbulencefig_2}(a) and \ref{turbulencefig_2}(b), we see a fuzzy distribution of an infinite number of singularity pairs in very short distances after the obstruction. By propagation to further distances, the productions of self-healing properties of the Bessel beam gradually emerge, and the distribution of singularities tends to take a regular form (see Fig. \ref{turbulencefig_2}(c)). Up to $2\,$cm,  as observed from Fig. \ref{turbulencefig_2}(d), we still see an odd number of singularities in the inner region, implying that the effects of the obstruction have not been removed yet. However, after around $10\,$ cm of propagation distance, the appearance of an extra singularity appears in the internal ring is evident from Fig. \ref{turbulencefig_2}(e) that restores the balance of the topological index existing in the initial beam confirming the self-healing consequences. This equity of positive and negative singularities is preserved in propagation to further distances as evident from Fig. \ref{turbulencefig_2}(f).\par

A similar phenomenon of creation of singularities has been demonstrated previously explained through the Hilbert Hotel mechanism \cite{Gbur:16} with this noticeable difference that in the Hilbert Hotel process, the evolution of singularities is accompanied by the generation and annihilation of singularity dipoles (pairs of opposite signs) representing a pair of room-guest in a hotel and the extra singularity appears through the shift of each guest to its next room. However, in the analyzed Poincar\'{e}-Bessel beam, the pairs that generate and annihilate along with the propagation of the beam are composed of singularities with similar index signs (room-room or guest-guest pairs) and no transition of singularities complying with the Hilbert Hotel definition is observed. Therefore, the creation of a new singularity resulting from the self-healing of the Bessel beam should have another origin that can be explored in the future.

\section{Experimental details}

\begin{figure}[t]
\centering
\includegraphics[width=\linewidth]{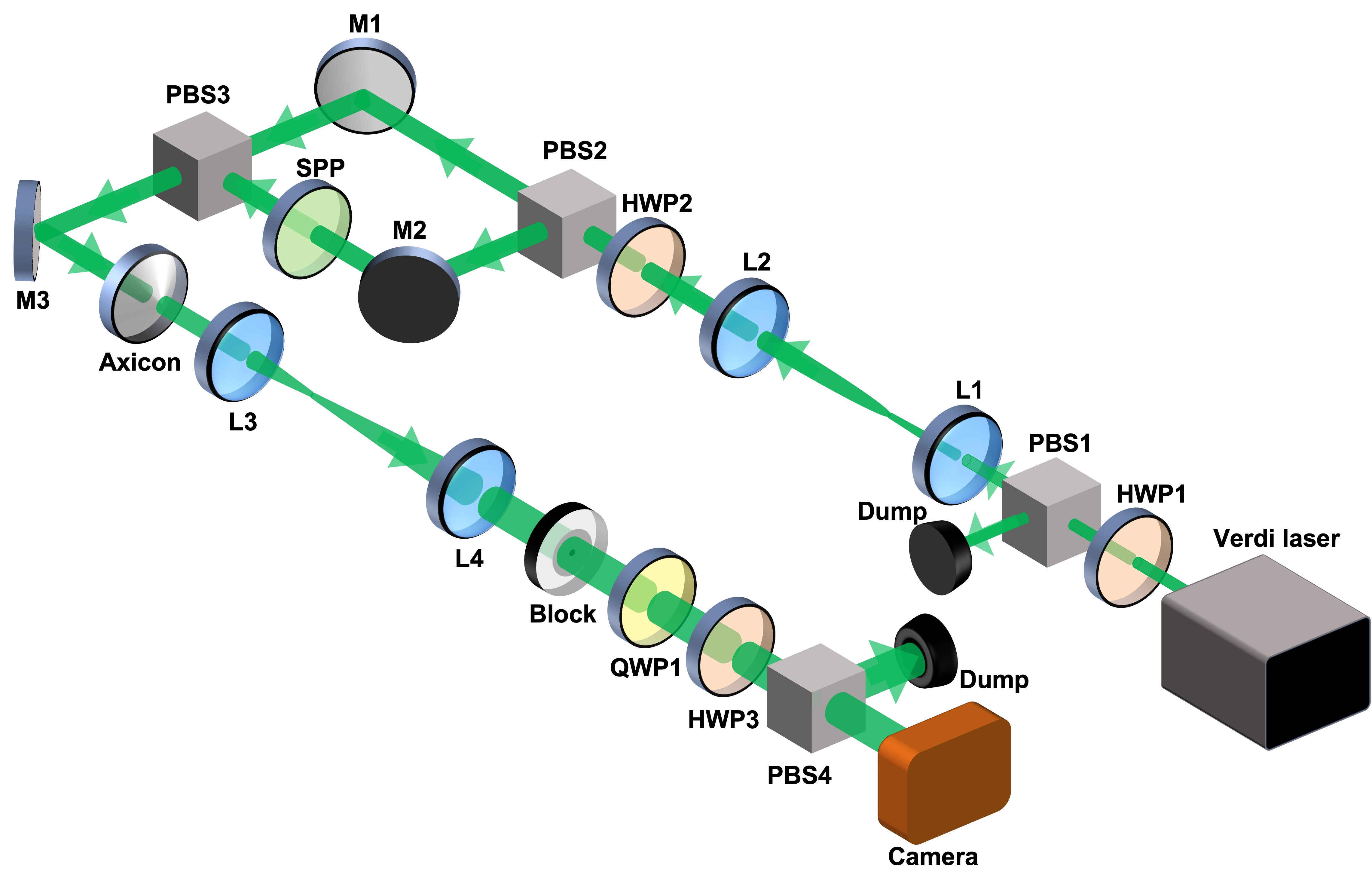}
\caption{Experimental setup for the generation of Poincar\'{e}-Bessel beam. HWP1-3: $\lambda/2$ plates; PBS1-4: polarizing beam splitter cubes;  SPP: spiral phase plate; QWP1: $\lambda/4$ plate; M1-3: mirrors; L1-4: lenses}
\label{setup}
\end{figure}

The schematic experimental scheme for the generation and study of full Poincar\'{e}-Bessel beam is shown in Fig. \ref{setup}.  A continuous wave (cw), single-frequency, green laser (Coherent, Verdi V10)  providing maximum output power of 10 W in $TEM_{00}$ spatial profile with $M^2<1.1$ at 532 nm is used as the primary laser source. Operating the laser at its maximum output power for reliable system performance, we have used a power attenuator comprised of the combination of $\lambda/2$ (HWP1) and a polarizing beam splitter (PBS1) cube to control the laser power to the experiment. A pair of plano-convex lenses, L1 and L2, of focal length, $f_1$ = 50 mm and $f_2$ = 200 mm, respectively, placed in 2$f_1$-2$f_2$ configuration is used to expand the laser beam. The $\lambda/2$ (HWP2) plate is used to control  the relative intensity between the two arms of the Mach-Zehnder interferometer (MZI) configured with two plane mirrors, M1 and M2, and two PBSs, PBS2, and PBS3. The full Poincar\'{e} beam is generated \cite{Beckley:10} by placing the spiral phase plate (SPP) in one of the arms (here between mirror, M2, and PBS3) of MZI. The SPP has the transverse thickness variation corresponding to the phase variation of the vortex order of $l$ = 1. As a result, the vertical polarized Gaussian beam of the reflected arm of MZI on propagation through the SPP and subsequent coaxial superposition with the horizontal polarized Gaussian beam of the transmitted arm on the PBS3 produces the full Poincar\'{e} beam with the electric field, 
$\alpha \ket{H,0}+\beta \ket{V,l}$. Here $\alpha$ and $\beta$ satisfying $\alpha^2 + \beta^2$ = 1, are the relative amplitudes of the orthogonal polarization modes of the FP beam. The $H$, and $V$ are the horizontal and vertical polarization states of the constituent beams, and $l$ is the order of the vortex beam. The values of $\alpha$ and $\beta$, can be controlled by varying the angle, $\theta$, of the HWP2 angle by $\alpha=\cos{\theta/2}$ and $\beta=\sin{\theta/2}$. The full Poincar\'{e} beam on propagation through the Axicon with an apex angle of $196^o$ is transformed into a Poincar\'{e}-Bessel beam. The Poincar\'{e}-Bessel beam then expanded with the second pair of plano-convex lenses, L3 and L4, of focal lengths of $f_3$ = 50 mm and $f_4$ = 300 mm, respectively in 2$f_3$-2$f_4$ imaging configuration. The polarization state of the beam is characterized using the standard Stokes measurement technique \cite{goldstein2017polarized} with the help of a quarter-wave plate (QWP), HWP3, PBS4, and the CCD camera. For the self-healing study, an obstacle (Block) made of a microscope cover slip with a black dot of diameter 0.2 mm at the center is used in the experiment.
\section{Results and discussion}
\subsection{Characteristics of Poincar\'{e}-Bessel beam}
First, we have characterized the polarization characteristics of the full Poincar\'{e}-Bessel beam generated through the mode transformation of the full Poincar\'{e} beam by the axicon. The electric field of the full Poincar\'{e} beam at the output of the MZI having SPP corresponding to vortex order, $l$, can be written as \cite{Beckley:10}
%
\begin{equation}\label{1}
    \begin{aligned}
  \mathbf{E} &=\alpha U^0_{G}(r,\phi)\mathbf{\hat{x}}+\beta  U^l_{LG}(r,\phi) \mathbf{\hat{y}}\\  
  &=U_0\left[\alpha \mathbf{\hat{x}}+\beta \left(\frac{\sqrt{2}r}{w}\right)^{|l|} LG_{l}\left(\frac{2r^2}{w^2}\right) e^{\pm il\phi} \mathbf{\hat{y}}\right]
  \end{aligned}
\end{equation}

%
with $\hat{x}$ and $\hat{y}$ representing horizontal and vertical polarization, respectively. $\alpha$  and $\beta$ are the relative weights of the two orthogonal polarized beams, and $\alpha^2 + \beta^2$ = 1. Here, $U^l_{LG}$, $U^0_{G}$ are the electric field amplitude of the vortex and Gaussian beams, respectively. Further, the Gaussian electric field amplitude at the origin is represented by $U_0$. The parameters $w$, $r$, and $\phi$ correspond to the beam waist, radial, and azimuthal components of the beam. On propagation through the axicon, the input full Poincar\'{e} beam represented by Eq. (\ref{1}) transformed into a Poincar\'{e}-Bessel beam; the superposition of orthogonal polarized Bessel beams of order same as the topological charge or order of the respective orthogonal polarized beams of the input full Poincar\'{e} beam. Therefore, the electric field of the Poincar\'{e}-Bessel beam can be represented as,

%
\begin{equation}\label{2}
    \begin{aligned}
     \mathbf{E} &=\alpha U^0_{J}(r,\phi) \mathbf{\hat{x}}+\beta U^l_{J}(r,\phi) \mathbf{\hat{y}}\\      
    &=U_0 \left[\alpha J_0\left(k_r r\right) \mathbf{\hat{x}} +\beta  J_l\left(k_r r\right) e^{\pm i l \phi}  \mathbf{\hat{y}}\right]
    \end{aligned}
\end{equation}
%
Here, $J_0\left(k_r r\right)$ and $J_l\left(k_r r\right)$ are $0^{th}$ and $l^{th}$ order Bessel beams, respectively. The maximum electric field amplitude at the origin is represented by the constant $U_0$. The radial wave vector $k_r$ can be written in terms of the wave vector($k$) and axicon apex angle($\gamma$) and refractive index ($n$) as $k_r=k(n-1)cos(\gamma)$.

%
\begin{figure}[t]
\centering
\includegraphics[width=\linewidth]{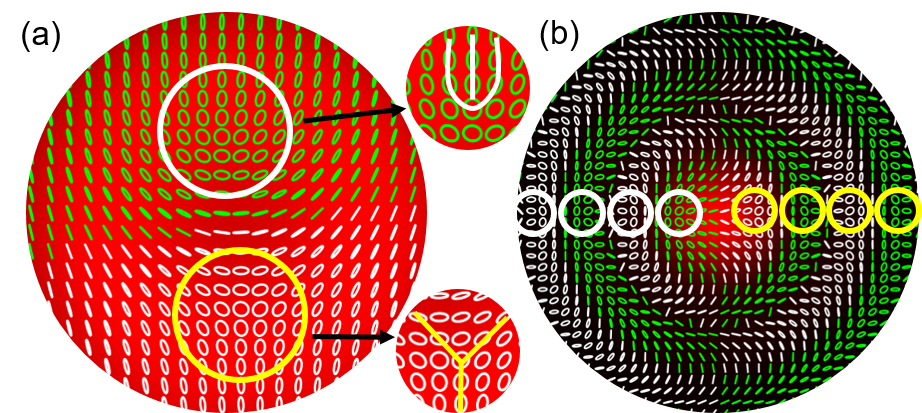}
\caption{Polarization distributions of (a) full Poincar\'{e} and (b) Poincar\'{e}-Bessel beams calculated using the experimental parameters in Eq. \ref{1} and Eq. \ref{2}. The insets are the magnified images of lemon and star polarization singularities marked by white and yellow circles.}
\label{theory}
\end{figure}
%

Using the experimental parameters (diameter (FWHM) of Gaussian and vortex beams as $\sim$3.4 mm and $\sim$6.8 mm, respectively, and their relative intensity weightage $\alpha/\beta$ = $1$) in Eq. (\ref{1}) and Eq. (\ref{2}) and the theoretically calculated Stokes parameters, we have calculated the orientation ($\psi$) and ellipticity ($\chi$) \cite{goldstein2017polarized} of the polarization ellipse of the input full Poincar\'{e} beam and corresponding Poincar\'{e}-Bessel beams. The results are shown in Fig. \ref{theory}. As evident from Fig. \ref{theory}(a), the transverse distribution of the polarization ellipse of the input beam of vortex order, $l$ =1, contains C-point polarization singularity in the form of pair of lemon (see the white circle) and star (see the yellow circle) and a single L-line confirming the vortex order of the full Poincar\'{e} beam to be $l$ = 1, same as our recent report \cite{Subith2023Coverage}. For clear observation, we have zoomed the section of polarization singularities as shown in the inset of Fig. \ref{theory}. Throughout the manuscript, we identify the white and yellow color circles as the lemon and star polarization singularities, respectively, if otherwise presented. However, the polarization distribution of the Poincar\'{e}-Bessel beam, as shown in Fig. \ref{theory}(b), shows a very interesting pattern where each concentric circle contains a pair of lemons (see the white circle) and stars (see the yellow circle) singularities. As the Bessel beam has characteristic intensity distribution of concentric rings and infinite spatial extend, we observe the Poincar\'{e}-Bessel beam to carry infinite pairs of lemon and star singularities. A careful observation of the polarization distribution indicates that each of the rings of the  Poincar\'{e}-Bessel beam contains a large number of polarization states, and the same polarization states are repeating in all the rings as if the Poincar\'{e}-Bessel beam consists of an infinite number of full Poincar\'{e} beams. 

\indent

\begin{figure}[t]
\centering
\includegraphics[width=\linewidth]{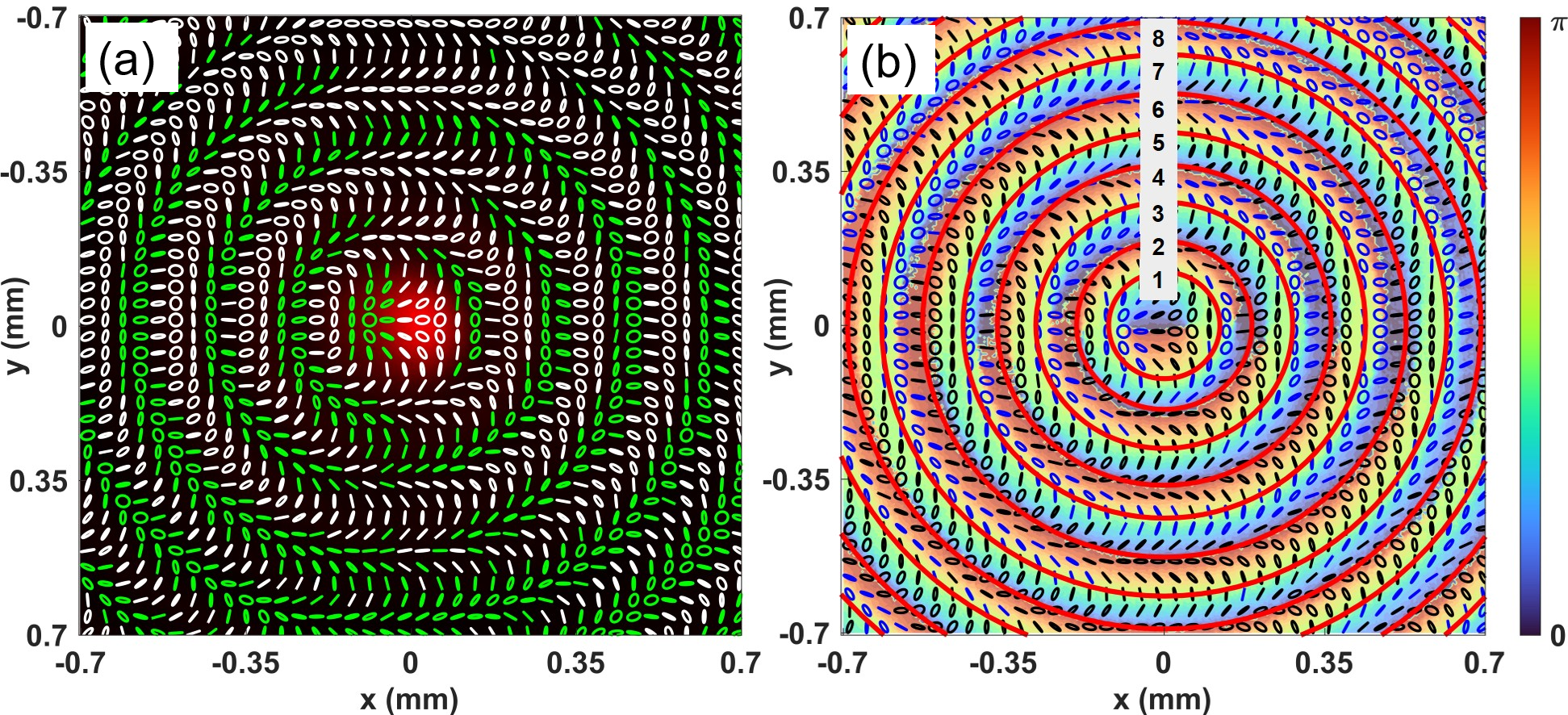}
\caption{Experimentally measured polarization distribution of Poincar\'{e}-Bessel beam with (a) beam intensity and (b) ellipse orientation map at the background. The red colour rings identify the characteristic rings of the Poincar\'{e}-Bessel beams. The rings are also marked integer numbers for further studies.}
\label{experimental_polarization_distribution}
\end{figure}

\begin{figure}[t]
\centering
\includegraphics[width=\linewidth]{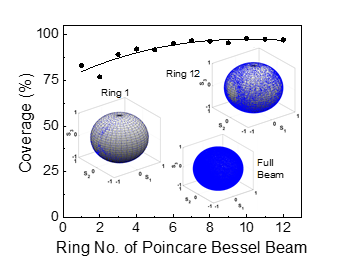}
\caption{The variation of polarization coverage of each ring of Poincar\'{e}-Bessel beam. Inset images show the polarization distribution of ring 1, ring 12, and the entire Poincar\'{e}-Bessel beam on the Poincar\'{e} sphere.}
\label{polarization_coverage}
\end{figure}

To confirm such interesting polarization characteristics of the experimentally generated Poincar\'{e}-Bessel beam, we recorded the intensity distribution for different polarization projections (using the combination of $\lambda$/4 and $\lambda$/2 plates at different combinations of the angles, the PBS). Using these intensity distributions, we have calculated the Stokes parameters, $S_1, S_2,$ and $S_3$, and measure the orientation ($\psi$) and ellipticity ($\chi$) of the polarization ellipse. The results are shown in Fig. \ref{experimental_polarization_distribution}.
As evident from Fig. \ref{experimental_polarization_distribution}(a), the generated Poincar\'{e}-Bessel beam has polarization distribution in close agreement with the theoretical results (see Fig. \ref{theory}(b)) with an infinite number of pairs lemon and star singularities (one pair in each ring of the Bessel beam containing an infinite number of concentric rings). As expected, the Poincar\'{e}-Bessel beam generated by the full Poincar\'{e} beam has central intensity maxima, and each ring contains a large number of polarization states. To get further perspective, we have recorded the polarization ellipse orientation (0 - $\pi$) of the Poincar\'{e}-Bessel beam along with the polarization distribution. The results are shown in Fig. \ref{experimental_polarization_distribution}(b). As evident from the Fig. \ref{experimental_polarization_distribution}(b), the ellipse orientation pattern of each ring (marked by red colour and identified by the numbers 1, 2, 3....) has two singular points with opposite polarization ellipse orientation (counterclockwise, star and clockwise, lemon directions) confirming the presence of pair of polarization singularities in each ring and an infinite number of polarization singularity pairs in the transverse spatial distribution of the Poincar\'{e}-Bessel beam.  
%
%
Recently, we have devised a new method to estimate the polarization coverage of the full Poincar\'{e} beam \cite{Subith2023Coverage}. Using the same treatment here, we have calculated the polarization coverage of each ring of the Poincar\'{e}-Bessel beam as marked in Fig. \ref{experimental_polarization_distribution}(b). The results are shown in Fig. \ref{polarization_coverage}. As evident from Fig. \ref{polarization_coverage}, the polarization coverage of the first ring of the Poincar\'{e}-Bessel beam is >75$\%$. However, there is an increase in polarization coverage with ring number, and finally, all rings have polarization coverage >97$\%$. Although we expect to have the same polarization coverage in all rings, however, relatively lower polarization coverage for the central ring of the Bessel beam arises from the experimental artifact limiting; the restriction in the number of useful camera pixels in the measurement process arising from the lower spatial extent of the central ring. The slightly lower polarization coverage for the second ring is due to the asymmetry in the central lobe of the first-order Bessel beam generated from the vortex. The artifacts gradually die out with the increase in camera pixel numbers to accommodate the increase in the spatial extent of the Bessel rings. Therefore, we observe the initial increase of polarization coverage, which finally saturates, resulting in the same value for all rings, as expected. For further support, we have calculated the area (number of pixels) varying from $4.8\times10^{-2} mm^2 $ (2019) to $53.7\times10^{-2} mm^2 $ (23633) for Ring number 1 to Ring no. 12 of the Poincar\'{e}-Bessel beam. However, as the coverage of more than 75$\%$ of the Poincar\'{e} sphere is deemed acceptable as a full Poincar\'{e} beam for many applications \cite{Arora:19}, we can safely say that each ring of the Poincar\'{e}-Bessel beam is a full Poincar\'{e} beam. We have also presented the polarization Poincar\'{e} sphere of each ring of Poincar\'{e}-Bessel beam as the inset of Fig. \ref{polarization_coverage}. As expected, the Poincar\'{e} sphere, as evident from the inset of Fig. \ref{polarization_coverage}, gets populated with the number of points for the Bessel beam rings away from the center without increasing the polarization coverage substantially. Finally, the polarization coverage of the entire Poincar\'{e}-Bessel beam, as also seen from the inset of Fig. \ref{polarization_coverage}, is around 100$\%$ and contains a large number of data points on the corresponding Poincar\'{e} sphere. Since the polarization states present in each ring of the Poincar\'{e}-Bessel beam covers the entire surface of the Poincar\'{e} sphere, one can imagine the polarization coverage of the whole Poincar\'{e}-Bessel beam as the superposition of an infinite number of same polarization states resulting the net polarization coverage same as the single ring. Such interesting property of Poincar\'{e}-Bessel beam supports the self-healing characteristics of polarization coverage, the same as the intensity self-healing of the Bessel beam.   
%
\subsection{Self healing of Poincar\'{e}-Bessel beam}
\begin{figure}[t]
\centering
\includegraphics[width=\linewidth]{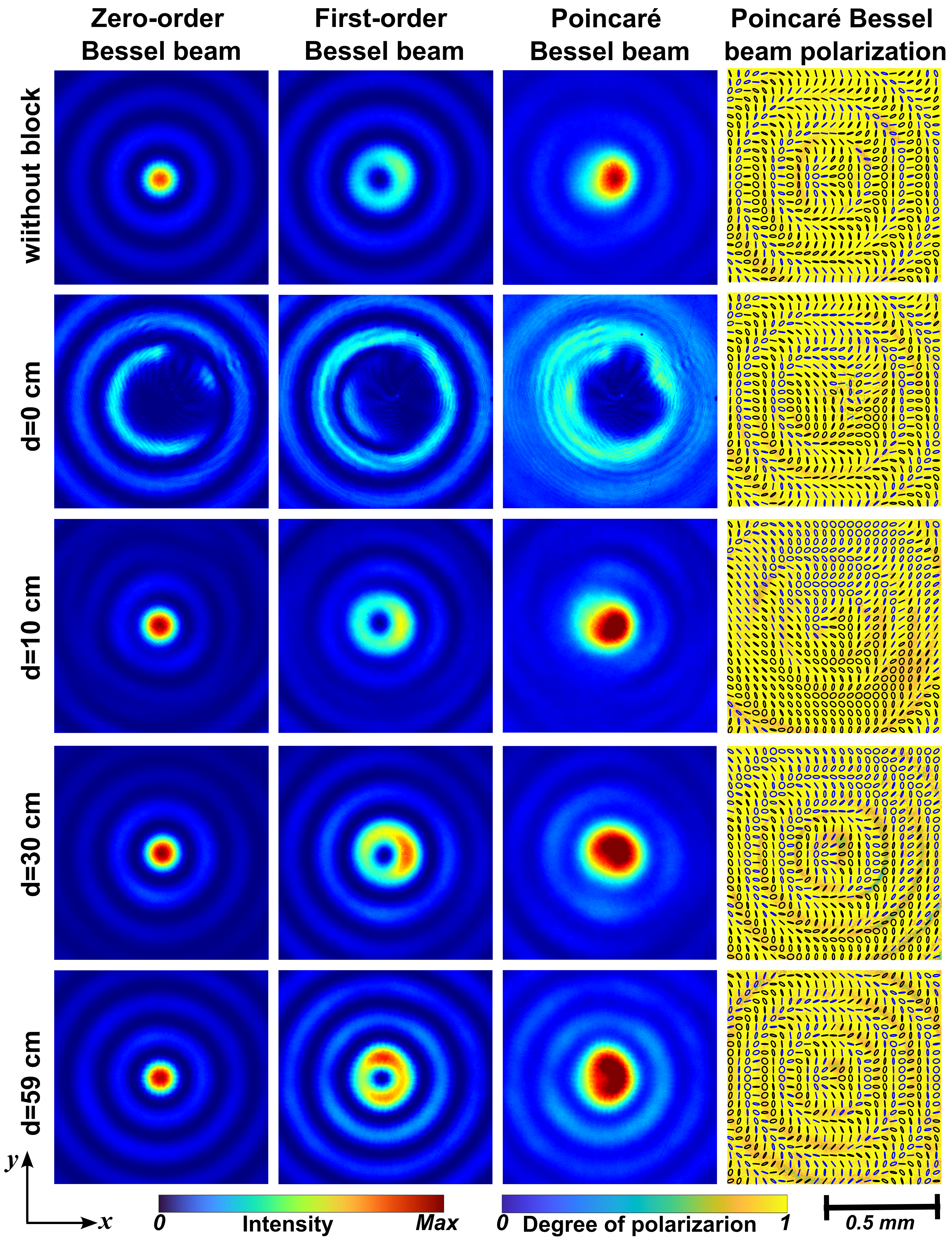}
\caption{Verification of self-healing characteristics of (first column) zero-order, (second column) first-order, and (third column) Poincar\'{e}-Bessel beam from the intensity distribution of the beam along propagation. Observation of (fourth column) polarization self-healing characteristics of the Poincar\'{e}-Bessel beam.}
\label{center_polarization_map}
\end{figure}

Knowing the polarization characteristics, we have studied the intensity and polarization self-healing properties of the Poincar\'{e}-Bessel beam. For the intensity self-healing study, we have recorded the intensity profile of the Bessel beam generated through the Gaussian, scalar vortex of order, $l$ = 1, and the full Poincar\'{e} beams as input to the axicon. On the other hand, we have estimated the degree of polarization, ellipse orientation, and ellipticity of the Poincar\'{e}-Bessel beam using the Stokes parameters calculated from the intensity profile of the Bessel beam recorded for different polarization projections. The results are shown in Fig. \ref{center_polarization_map}. As expected, the zero-order Bessel beam, first-order Bessel beams, and Poincar\'{e}-Bessel beam, shown by the first, second, and third columns of Fig. \ref{center_polarization_map}, respectively, have disturbed intensity distribution at the beam center resulting from the beam abstraction. However, all the beams start regaining and maintaining their initial spatial intensity distribution after a propagation distance of d = 10 cm with complete healing at d = 59 cm. 
%
%

To show the variation of the degree of polarization of the Poincar\'{e}-Bessel beam during the self-healing process, we have used the colour map with blue and yellow colours representing unpolarized and perfectly polarized beams, respectively. The fourth column of Fig. \ref{center_polarization_map} shows the polarization distribution in combination with the degree of polarization in the background. As evident from the fourth column of Fig. \ref{center_polarization_map}, the abstraction, although it disturbs the polarization distribution of the beam, has negligible or no impact on the degree of polarization. The Poincar\'{e}-Bessel beam maintains a high degree of polarisation ($\approx$1) throughout its cross-section. On the other hand, the concentric ring profile of the polarization distribution follows the same self-healing characteristics as the beam's intensity profile with propagation.

\begin{figure}[h]
\centering
\includegraphics[width=\linewidth]{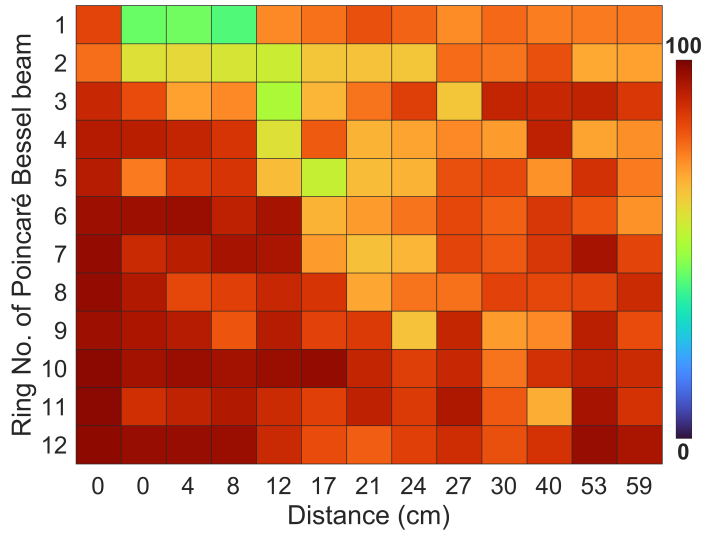}
\caption{Variation of polarization coverage of different rings of the Poincar\'{e}-Bessel beam along propagation during the self-healing process. The lower polarization coverage due to the beam obstruction is gradually moving in the outward direction during the self-healing process.}
\label{pol_cove_along_propagation}
\end{figure}

\begin{figure}[t]
\centering
\includegraphics[width=\linewidth]{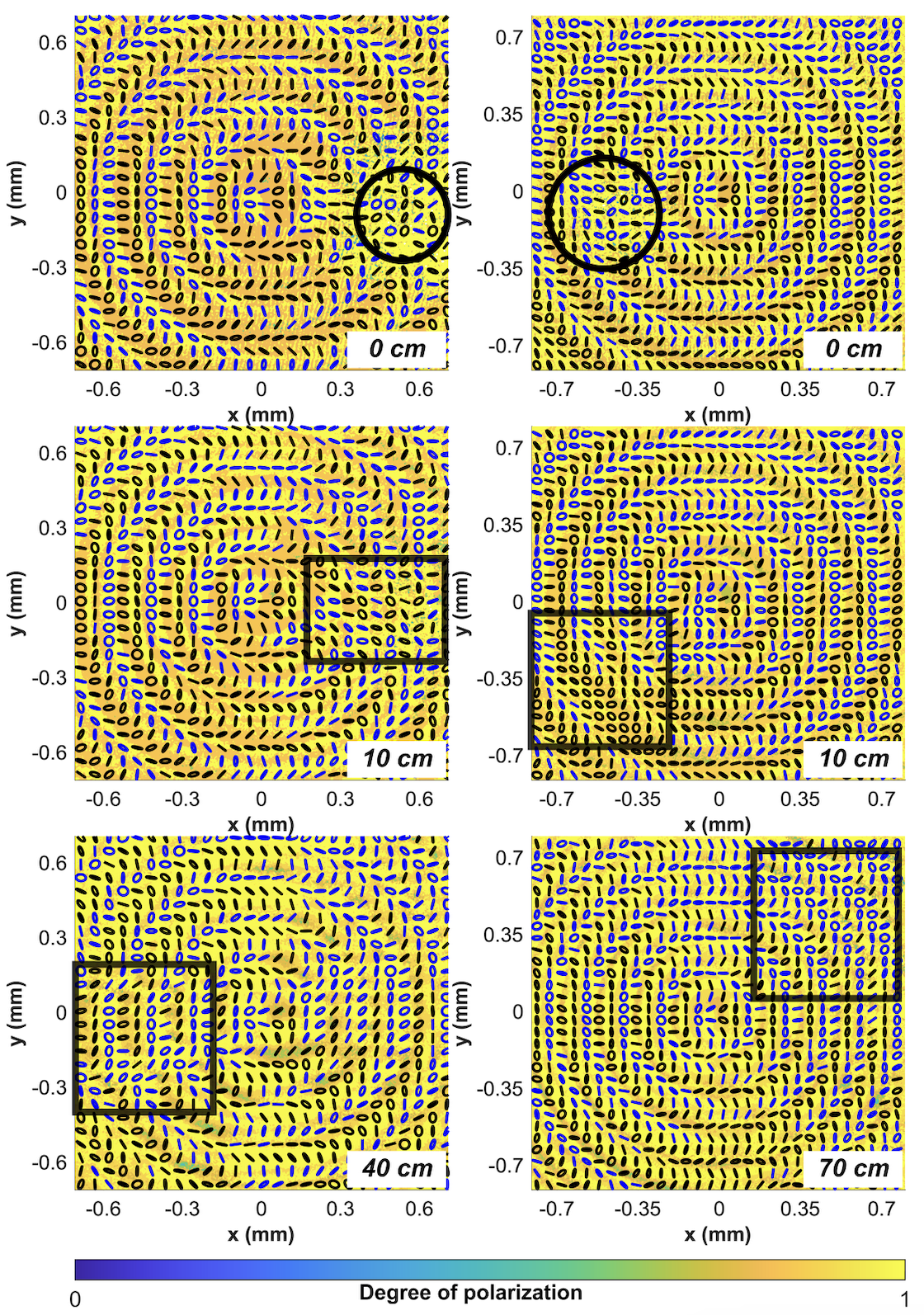}
\caption{Dynamics of polarization distribution and degree of polarization of Poincar\'{e}-Bessel beam along propagation after the beam obstruction by two blocks of different sizes and positions. Black circles and squares mark the position of polarization disturbance for easy identification. }
\label{dif_block}
\end{figure}

To understand further the polarization self-healing characteristics of the Poincar\'{e}-Bessel beam, we have estimated the polarization coverage of each ring along propagation distance. The results are shown by colour chart in Fig. \ref{pol_cove_along_propagation}. The rows and columns of Fig. \ref{pol_cove_along_propagation} represent the ring number of propagation distance, respectively. It is evident from Fig. \ref{pol_cove_along_propagation} all the rings before the beam obstruction carry polarization coverage >84$\%$, same as Fig. \ref{polarization_coverage}. As expected, the polarization coverage of the central ring is low due to beam obstruction, while the polarization coverage of other rings remains unaffected. However, it is interesting to observe that the disturbance in the polarization coverage of the beam gradually travels away from the inner ring to the outer rings along the beam propagation and finally regains high polarization coverage for all rings. This observation confirms that the self-healing of the polarization coverage of the Poincar\'{e}-Bessel beam occurs due to the energy flow of the Bessel beam from the outward rings to the inside rings.        
%
We also blocked the beam at various positions on the transverse plane and calculated the degree of polarization and polarization distribution to understand the impact of beam block sites on the self-healing process. We have selected obstructions of two sizes (circles of diameter 0.7 mm and 0.5 mm) and positions. The results are shown in Fig. \ref{dif_block}. As expected, in both cases, the degree of polarization, as shown by the background color map to the polarization distribution of the Poincar\'{e}-Bessel beam in Fig. \ref{dif_block}, is unperturbed to the beam obstruction.  
On the other hand, it is observed from the first column of Fig. \ref{dif_block} that the beam block, as identified by the black circle, disrupts the polarization distribution initially at the locations it is placed. However, with beam propagation, it is observed from the second row of Fig. \ref{dif_block} that the disturbance gradually spreads away from the disturbed site. With further beam propagation, it is observed from the third row of Fig. \ref{dif_block} that the disturbance appears on the diametrically opposite side of the beam before resetting the effect beam block and returning to the initial polarization distribution. Like the intensity self-healing of the Bessel beams, the current observation clearly indicates that beam block size (much smaller than the beam size) and position does not have a detrimental effect on the intensity and polarization self-healing characteristics of the Poincar\'{e}-Bessel beam. 
%
%
%
\subsection{Polarization singularities}
\begin{figure}[h]
\centering
\includegraphics[width=\linewidth]{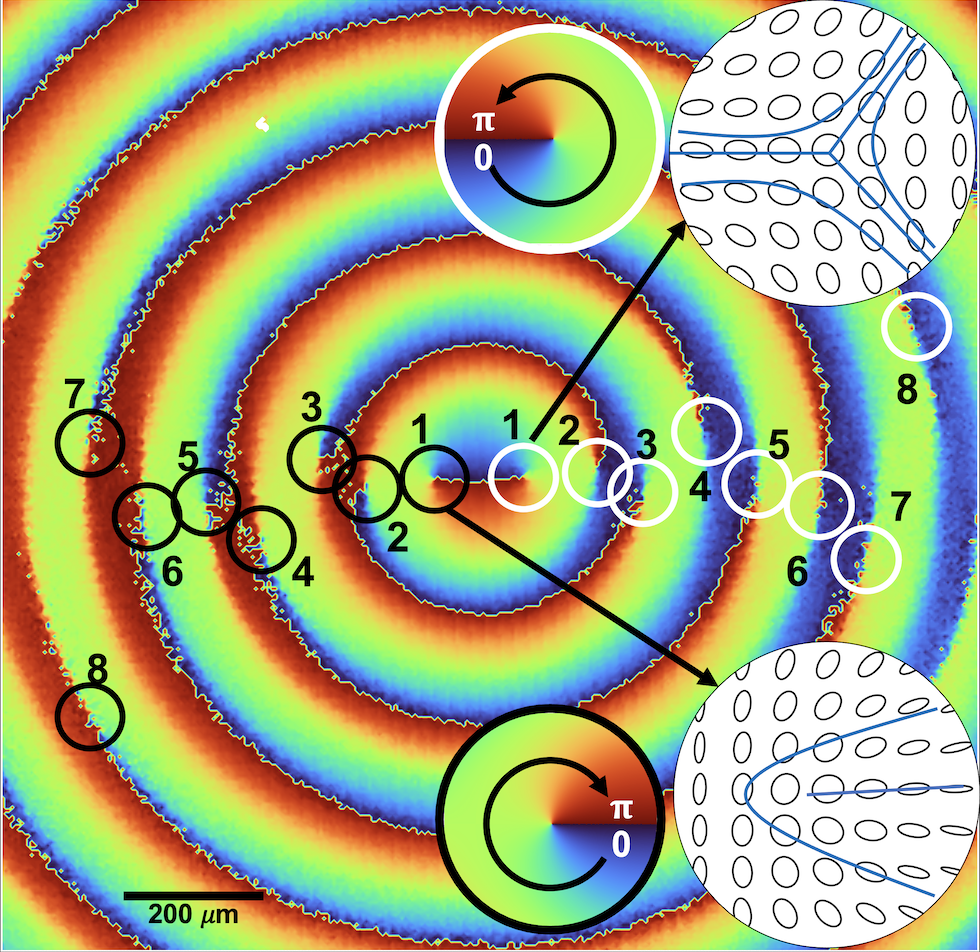}
\caption{Polarization ellipse orientation map of the Poincar\'{e}-Bessel beam showing the infinite series of C-point singularity pairs. The white and black circles identify the star (polarisation ellipse varying from $0-\pi$ in the counterclockwise direction) and lemon (polarisation ellipse varying from $0-\pi$ in the clockwise direction) singularities, respectively. The insets show the polarization distribution and corresponding polarisation ellipse orientation at C-point singularities.}
\label{ellipse_oriantation}
\end{figure}

Knowing the complete characteristics of the Poincar\'{e}-Bessel beam in different degrees of freedom, including the intensity, polarization distribution, and degree of polarization, we have studied the effect of the C-point singularity of the beam in the self-healing process. As presented in Fig. \ref{experimental_polarization_distribution}(a), the Poincar\'{e}-Bessel beam carries an infinite number of pairs of lemon and star singularities \textcolor{black}{in close agreement with the theory (see Fig. \ref{turbulencefig_1})}. To appreciate the observation and identify the C-points singularities, we have calculated the orientation of the polarisation ellipse varying from $0-\pi$ with the results shown in Fig. \ref{ellipse_oriantation}. For easy comprehension, we have identified the direction of ellipse orientation about the singularity point by $0-\pi$ in the counterclockwise direction with the white circle representing star singularity and the clockwise direction with the black circle representing lemon singularity. 
%
%
As expected, we observe the polarization ellipse orientation map of the Poincar\'{e}-Bessel beam in the rectangular basis to contain a series of lemon and star singularity pairs in each ring. As the ideal Bessel beams have infinite spatial extend and thus an infinite number of rings, we can clearly confirm the generation of infinite series of lemon and star polarization singularity pairs by transforming the full Poincar\'{e} beam into Poincar\'{e}-Bessel beam. The number of series is decided by the number of polarization singularity pairs of the input full Poincar\'{e} beam or simply the order of the vortex. Using higher order full Poincar\'{e} beam of vortex orders, $l$ = 2, 3  and 4, we have observed the resultant Poincar\'{e}-Bessel beam to carry 2, 3 and 4 infinite series of lemon and star polarization singularity pairs.  

%
\begin{figure}[t]
\centering
\includegraphics[width=\linewidth]{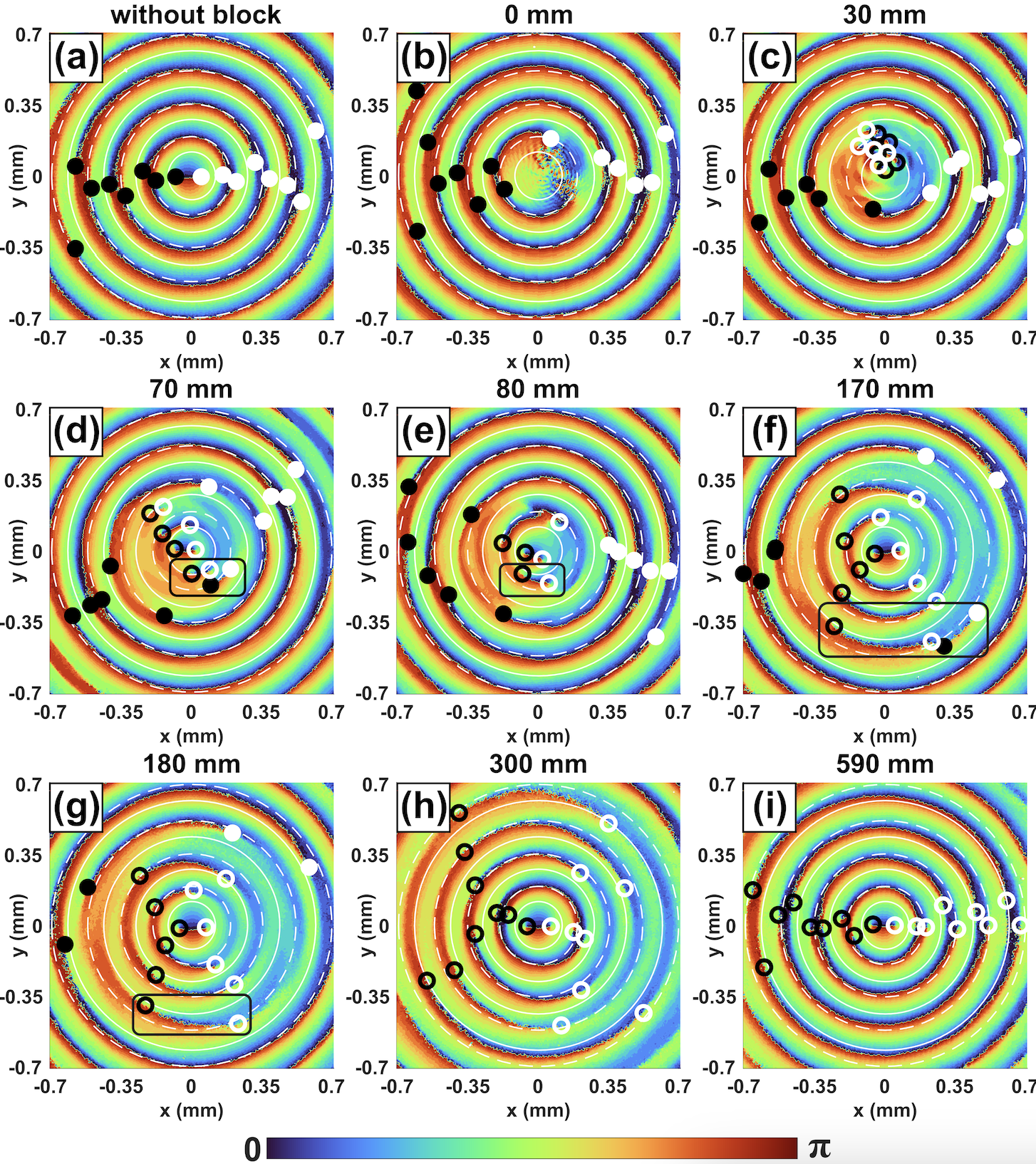}
\caption{Observation of dynamics of the infinite number of C-point singularity pairs of the Poincar\'{e}-Bessel beam along propagation after the beam obstruction. The star and lemon C-point singularities of the initial beam are marked by white and black dots, respectively. The newly formed star and lemon singularity pairs due to beam obstruction are identified by white and black circles. The black rectangles mark the annihilation of the old singularity with the new singularity.}
\label{C-points}
\end{figure}
%
Further, we have studied the dynamics of the c-point singularity of the Poincar\'{e}-Bessel beam in the self-healing process. Using the intensity distribution of the beam for different projections along beam propagation, we have derived the polarization ellipse orientation map with the results shown in Fig. \ref{C-points}. As expected, the Poincar\'{e}-Bessel beam contained an infinite series of star and lemon polarization singularities identified by the black and white dots in Fig. \ref{C-points}(a). As the star and lemon polarization, singularities have singularity indices, $I_c$ = +1/2 and $I_c$ = -1/2, respectively, the net topological charge of the polarization singularities in the Poincar\'{e}-Bessel beam can be considered to be zero. In the self-healing study, we purposefully adjusted the block position so that it created an asymmetry in the total number of star and lemon singularities. As evident from Fig. \ref{C-points}(b), the block has removed two stars and one lemon singularity from the polarization ellipse orientation map at the center, confirming the presence of asymmetry in the number of singularities right after the beam block. However, along beam propagation, as shown by Fig. \ref{C-points}(c-i), we observe exciting dynamics of C-point singularities. The beam block-induced perturbation results in the production of an infinite number of new C-point singularity pairs, lemon, and star, marked by black and white open circles (see Fig. \ref{C-points}(c)). Due to the limited spatial resolution arising from the beam size and the pixel size of the CCD, we have marked a limited number of new C-point singularity pairs and observed their propagation dynamics. It is interesting to observe that the newly generated singularity pairs (black and white open circles) annihilate with each other and also with the intrinsic singularity pairs (black and white dots) with beam propagation (see Fig. \ref{C-points} (d-i) and eventually take over the original singularity pairs in the center (Fig. \ref{C-points}(i)). After the self-healing process, it is interesting to observe that the Poincar\'{e}-Bessel beam has an infinite number of C-point singularity pairs resulting in the net topological charge of zero, the same as the initial beam. However, after further beam propagation, the self-healed Poincar\'{e}-Bessel beam, as shown by Fig. \ref{C-points}(i), regains the  position of the C-point singularity and reproduces the polarization ellipse orientation map, the same as the initial beam (see Fig. \ref{C-points}(a)). A close look at the dynamics of the Poincar\'{e}-Bessel beam in the self-healing process, especially the polarization ellipse orientation map of Fig. \ref{C-points}(a), (b) and (i), we see the existence of an optical analogy to the mathematics of transfinite numbers through the Hilbert's Hotel like setting. To elaborate further, let's consider the polarization ellipse orientation map as Hilbert's hotel, where the star singularity (black dots) and lemon singularity (white dots)  represent the rooms and guests, respectively. From Fig. \ref{C-points}(a), it is evident that Hilbert's hotel, having an infinite number of rooms, is fully occupied by the guest, as the Poincar\'{e}-Bessel beam carries an infinite number of C-point singularity pairs. Using the beam block, we have created a situation (see Fig. \ref{C-points}(b)) where the number of lemon singularities (white dot) is more than one than the number of star singularities (black dots). This situation can be considered as if the hotel is fully occupied, but an extra guest (white dot) has appeared for accommodation. Although we don't see the exact transition as described in the famous lecture of the mathematician David Hilbert and demonstrated in optics \cite{Gbur:16, Wang:17, Chen:22, Kumar:23} during the self-healing process of the Poincar\'{e}-Bessel beam, however, at the end of the self-healing process we do observe that the additional guest has been accommodated in Hilbert's hotel, making the hotel again fully occupied with a number of rooms (star singularity) is equal to the number of guests (lemon singularity). However, during the self-healing process, we see exciting features that can be used as optical analogies to understand the rich mathematics of transfinite numbers, which is beyond the scope of the current study and might lead to new submissions. \textcolor{black}{The self-healing dynamics of the experimentally measured Poincar\'{e}-Bessel beam closely match with the theoretical results presented in Fig. \ref{turbulencefig_2}.} 
\section{Conclusion}

In conclusion, we have experimentally studied the polarization characteristics of the Poincar\'{e}-Bessel beam in close agreement with the theoretical results. The use of a Poincar\'{e} beam in a rectangular basis containing all polarization states covered by the surface of the Poincar\'{e} sphere produces Poincar\'{e}-Bessel beam with each ring having polarization coverage >75$\%$. Again the polarization coverage of the Poincar\'{e}-Bessel beam is the same or slightly higher than the polarization coverage of any of the rings. Therefore, one can consider the Poincar\'{e}-Bessel beam as the superposition of an infinite number of Poincar\'{e} beams. Further, it is observed that the polarization structure of the Poincar\'{e}-Bessel beam shows self-healing characteristics like intensity self-healing after being abstracted. We also observed the degree of polarization of the beam has no impact on the beam obstruction. Using the polarization ellipse orientation map, we observe the beam to carry an infinite number of C-point singularity (lemon and star) pairs. The number of such infinite series is decided by the number of C-point singularity (lemon and star) pairs present in the input full Poincar\'{e} beam. As the number of C-point singularity pairs of the full Poincar\'{e} beam is equal to the vortex order of the constituent superposed orthogonal polarized beams, one can generate any number infinite series of C-point singularity (lemon and star) pairs by simply adjusting the vortex order. Using the single series of infinite numbers of  C-point singularity (lemon and star) pairs, we transition the dynamics of the C-point singularities in the self-healing process and observe Hilbert's hotel-like process addressing the connection to the mathematics of infinite sets. The current study can, in principle, be used for imaging objects even in the presence of depolarizing surroundings, studying turbulent atmospheric channels for communication and rich mathematical concepts of transfinite numbers.
\section*{AUTHOR DECLARATIONS}
\subsection*{Conflict of Interest}
The authors have no conflicts to disclose.
\subsection*{Author Contributions}
S. K., and A. P. developed the experimental setup and performed measurements, data analysis, and numerical simulation. A. S. and G. G. lead the theoretical study, derived analytical formulas, and numerical simulation. G. S. developed the ideas and lead the project. All authors participated in the discussion and contributed to the manuscript writing.

%
%
\section*{DATA AVAILABILITY}
The data that support the findings of this study are available from the corresponding author upon reasonable request.

\bibliography{references}

\begin{thebibliography}{10}
\newcommand{\enquote}[1]{``#1''}

\bibitem{Otte2020OpticalManipulation}
E.~Otte and C.~Denz, \enquote{{Optical trapping gets structure: Structured
  light for advanced optical manipulation},} {\protect\JournalTitle{Applied
  Physics Reviews}} \textbf{7}, 041308 (2020).

\bibitem{Wang2012OpticalBeams}
L.-G. Wang, \enquote{{Optical forces on submicron particles induced by full
  Poincar{\'{e}} beams},} {\protect\JournalTitle{Optics Express, Vol. 20, Issue
  19, pp. 20814-20826}} \textbf{20}, 20814--20826 (2012).

\bibitem{liu2018nonlinear}
H.~Liu, H.~Li, Y.~Zheng, and X.~Chen, \enquote{{Nonlinear frequency conversion
  and manipulation of vector beams},} {\protect\JournalTitle{Optics Letters}}
  \textbf{43}, 5981--5984 (2018).

\bibitem{Toyoda2013TransferNanostructures}
K.~Toyoda, F.~Takahashi, S.~Takizawa, Y.~Tokizane, K.~Miyamoto, R.~Morita, and
  T.~Omatsu, \enquote{{Transfer of Light Helicity to Nanostructures},}
  {\protect\JournalTitle{Physical Review Letters}} \textbf{110}, 143603 (2013).

\bibitem{Omatsu2019ALight}
T.~Omatsu, K.~Miyamoto, K.~Toyoda, R.~Morita, Y.~Arita, and K.~Dholakia,
  \enquote{{A New Twist for Materials Science: The Formation of Chiral
  Structures Using the Angular Momentum of Light},}
  {\protect\JournalTitle{Advanced Optical Materials}} \textbf{7}, 1801672
  (2019).

\bibitem{Chen2013ImagingSystem}
R.~Chen, K.~Agarwal, C.~J.~R. Sheppard, and X.~Chen, \enquote{{Imaging using
  cylindrical vector beams in a high-numerical-aperture microscopy system},}
  {\protect\JournalTitle{Optics Letters}} \textbf{38}, 3111--3114 (2013).

\bibitem{Torok2004TheMicroscopy}
P.~T{\"{o}}r{\"{o}}k and P.~Munro, \enquote{{The use of Gauss–Laguerre vector
  beams in STED microscopy},} {\protect\JournalTitle{Opt. Express.}}
  \textbf{12} (2004).

\bibitem{Ndagano2018CreationCommunication}
B.~Ndagano, I.~Nape, M.~A. Cox, C.~Rosales-Guzman, and A.~Forbes,
  \enquote{{Creation and Detection of Vector Vortex Modes for Classical and
  Quantum Communication},} {\protect\JournalTitle{Journal of Lightwave
  Technology}} \textbf{36}, 292--301 (2018).

\bibitem{Lei2017ApproachBeams}
T.~Lei, Z.~Wu, S.~Gao, W.~Qiao, Z.~Li, and X.~Yuan, \enquote{{Approach to
  multiplexing fiber communication with cylindrical vector beams},}
  {\protect\JournalTitle{Optics Letters, Vol. 42, Issue 13, pp. 2579-2582}}
  \textbf{42}, 2579--2582 (2017).

\bibitem{Gbur2017SingularOptics}
G.~J. Gbur, \emph{{Singular Optics}} (CRC Press, 2017).

\bibitem{Gu:09}
Y.~Gu, O.~Korotkova, and G.~Gbur, \enquote{Scintillation of nonuniformly
  polarized beams in atmospheric turbulence,} {\protect\JournalTitle{Opt.
  Lett.}} \textbf{34}, 2261--2263 (2009).

\bibitem{Wei:15}
C.~Wei, D.~Wu, C.~Liang, F.~Wang, and Y.~Cai, \enquote{Experimental
  verification of significant reduction of turbulence-induced scintillation in
  a full poincar{\'e} beam,} {\protect\JournalTitle{Opt. Express}} \textbf{23},
  24331--24341 (2015).

\bibitem{Kumar:23}
S.~Kumar, A.~Ghosh, C.~Kaushik, A.~Shiri, G.~Gbur, S.~Sharma, and G.~Samanta,
  \enquote{Simple experimental realization of optical hilbert hotel using
  scalar and vector fractional vortex beams,} {\protect\JournalTitle{arXiv
  preprint arXiv:2303.11007}}  (2023).

\bibitem{Khonina2020BesselReview}
S.~N. Khonina, N.~L. Kazanskiy, S.~V. Karpeev, and M.~A. Butt, \enquote{{Bessel
  Beam: Significance and Applications—A Progressive Review},}
  {\protect\JournalTitle{Micromachines 2020, Vol. 11, Page 997}} \textbf{11},
  997 (2020).

\bibitem{mcgloin2005bessel}
D.~McGloin and K.~Dholakia, \enquote{Bessel beams: diffraction in a new light,}
  {\protect\JournalTitle{Contemporary physics}} \textbf{46}, 15--28 (2005).

\bibitem{stoian2018ultrafast}
R.~Stoian, M.~K. Bhuyan, G.~Zhang, G.~Cheng, R.~Meyer, and F.~Courvoisier,
  \enquote{Ultrafast bessel beams: advanced tools for laser materials
  processing,} {\protect\JournalTitle{Advanced Optical Technologies}}
  \textbf{7}, 165--174 (2018).

\bibitem{liu2020simultaneous}
Z.~Liu, X.~Tang, Y.~Zhang, Y.~Zhang, L.~Ma, M.~Zhang, X.~Yang, J.~Zhang,
  J.~Yang, and L.~Yuan, \enquote{Simultaneous trapping of low-index and
  high-index microparticles using a single optical fiber bessel beam,}
  {\protect\JournalTitle{Optics and lasers in engineering}} \textbf{131},
  106119 (2020).

\bibitem{dudutis2016non}
J.~Dudutis, P.~Ge{\v{C}}ys, and G.~Ra{\v{C}}iukaitis, \enquote{Non-ideal
  axicon-generated bessel beam application for intra-volume glass
  modification,} {\protect\JournalTitle{Optics express}} \textbf{24},
  28433--28443 (2016).

\bibitem{ambrosio2011integral}
L.~A. Ambrosio and H.~E. Hern{\'a}ndez-Figueroa, \enquote{Integral localized
  approximation description of ordinary bessel beams and application to optical
  trapping forces,} {\protect\JournalTitle{Biomedical optics express}}
  \textbf{2}, 1893--1906 (2011).

\bibitem{glukhova2022vector}
S.~A. Glukhova and M.~A. Yurkin, \enquote{Vector bessel beams: General
  classification and scattering simulations,} {\protect\JournalTitle{Physical
  Review A}} \textbf{106}, 033508 (2022).

\bibitem{bouchal1995non}
Z.~Bouchal and M.~Oliv{\'\i}k, \enquote{Non-diffractive vector bessel beams,}
  {\protect\JournalTitle{Journal of Modern Optics}} \textbf{42}, 1555--1566
  (1995).

\bibitem{PhysRevA.71.033411}
R.~J\'auregui and S.~Hacyan, \enquote{Quantum-mechanical properties of bessel
  beams,} {\protect\JournalTitle{Phys. Rev. A}} \textbf{71}, 033411 (2005).

\bibitem{mclaren2014self}
M.~McLaren, T.~Mhlanga, M.~J. Padgett, F.~S. Roux, and A.~Forbes,
  \enquote{Self-healing of quantum entanglement after an obstruction,}
  {\protect\JournalTitle{Nature communications}} \textbf{5}, 3248 (2014).

\bibitem{fahrbach2010microscopy}
F.~O. Fahrbach, P.~Simon, and A.~Rohrbach, \enquote{Microscopy with
  self-reconstructing beams,} {\protect\JournalTitle{Nature photonics}}
  \textbf{4}, 780--785 (2010).

\bibitem{planchon2011rapid}
T.~A. Planchon, L.~Gao, D.~E. Milkie, M.~W. Davidson, J.~A. Galbraith, C.~G.
  Galbraith, and E.~Betzig, \enquote{Rapid three-dimensional isotropic imaging
  of living cells using bessel beam plane illumination,}
  {\protect\JournalTitle{Nature methods}} \textbf{8}, 417--423 (2011).

\bibitem{cheng2016propagation}
M.~Cheng, L.~Guo, J.~Li, and Q.~Huang, \enquote{Propagation properties of an
  optical vortex carried by a bessel--gaussian beam in anisotropic turbulence,}
  {\protect\JournalTitle{JOSA A}} \textbf{33}, 1442--1450 (2016).

\bibitem{Shvedov:15}
V.~Shvedov, P.~Karpinski, Y.~Sheng, X.~Chen, W.~Zhu, W.~Krolikowski, and
  C.~Hnatovsky, \enquote{Visualizing polarization singularities in
  bessel-poincar\'{e} beams,} {\protect\JournalTitle{Opt. Express}}
  \textbf{23}, 12444--12453 (2015).

\bibitem{Holmes:19}
B.~M. Holmes and E.~J. Galvez, \enquote{Poincaré bessel beams: structure and
  propagation,} {\protect\JournalTitle{Journal of Optics}} \textbf{21}, 104001
  (2019).

\bibitem{ggs:tcps:1852}
G.~Stokes, \enquote{On the composition and resolution of streams of polarized
  light from different sources,} {\protect\JournalTitle{Trans. Camb. Phil.
  Soc.}} \textbf{9}, 399--416 (1852).

\bibitem{Beckley:10}
A.~M. Beckley, T.~G. Brown, and M.~A. Alonso, \enquote{Full poincar\'{e}
  beams,} {\protect\JournalTitle{Opt. Express}} \textbf{18}, 10777--10785
  (2010).

\bibitem{jd:josaa:1987}
J.~Durnin, \enquote{Exact solutions for nondiffracting beams. {I}. {T}he scalar
  theory,} {\protect\JournalTitle{J. Opt. Soc. Am. A}} \textbf{4}, 651--654
  (1987).

\bibitem{nye1974dislocations}
J.~F. Nye and M.~V. Berry, \enquote{Dislocations in wave trains,}
  {\protect\JournalTitle{Proceedings of the Royal Society of London. A.
  Mathematical and Physical Sciences}} \textbf{336}, 165--190 (1974).

\bibitem{Gbur:16}
G.~Gbur, \enquote{Fractional vortex hilbert's hotel,}
  {\protect\JournalTitle{Optica}} \textbf{3}, 222--225 (2016).

\bibitem{goldstein2017polarized}
D.~H. Goldstein, \emph{{Polarized light}} (CRC press, 2017), second edi ed.

\bibitem{Subith2023Coverage}
S.~Kumar, R.~K. Saripalli, A.~Ghosh, W.~T. Buono, A.~Forbes, and G.~Samanta,
  \enquote{Controlling the coverage of full poincar\'e beams through
  second-harmonic generation,} {\protect\JournalTitle{Phys. Rev. Appl.}}
  \textbf{19}, 034082 (2023).

\bibitem{Arora:19}
G.~Arora, Ruchi, and P.~Senthilkumaran, \enquote{Full poincar\'{e} beam with
  all the stokes vortices,} {\protect\JournalTitle{Opt. Lett.}} \textbf{44},
  5638--5641 (2019).

\bibitem{Wang:17}
Y.~Wang and G.~Gbur, \enquote{Hilbert's hotel in polarization singularities,}
  {\protect\JournalTitle{Opt. Lett.}} \textbf{42}, 5154--5157 (2017).

\bibitem{Chen:22}
X.~Chen, S.~Wang, C.~You, O.~S. Maga\~na Loaiza, and R.-B. Jin,
  \enquote{Experimental implementation of the fractional-vortex hilbert hotel,}
  {\protect\JournalTitle{Phys. Rev. A}} \textbf{106}, 033521 (2022).

\end{thebibliography}
\bibliographyfullrefs{references}

\end{document}